\documentclass[twocolumn,showpacs,superscriptaddress,amssymb,10pt]{revtex4}

\usepackage{graphicx}
\usepackage{dcolumn}
\usepackage{bm}
\usepackage{epsfig}
\usepackage{color}
\usepackage{longtable}

\definecolor{aogreen}{rgb}{0.0, 0.5, 0.0}
\def\ket#1{ $ \left\vert  #1   \right\rangle $ }
\def\ketm#1{  \left\vert  #1   \right\rangle   }

\def\sprm#1#2{  \left\langle #1 \left\vert \right. #2 \right\rangle   }

\def\mem#1#2#3{  \left\langle #1 \left\vert  #2 \right\vert #3 \right\rangle   }

\def\redmem#1#2#3{  \left\langle #1 \left\Vert
                  #2 \right\Vert #3 \right\rangle   }
\def\twobytwo#1#2#3#4{  \left( \begin{array}{cc}
                                   #1 & #2   \\[0.2cm]
                                   #3 & #4   \end{array} \right)   }

\begin{document}

\preprint{}
%
\title{Level sequence and splitting identification of closely-spaced energy levels by angle-resolved analysis
       of the fluorescence light}

\author{Z.~W.~Wu}
\affiliation{Helmholtz-Institut Jena, Fr\"o{}belstieg 3, D-07743 Jena, Germany}%
\affiliation{Key Laboratory of Atomic and Molecular Physics $\&$ Functional Materials of Gansu Province,
             College of Physics and Electronic Engineering, Northwest Normal University, Lanzhou 730070, P.R. China}

\author{A.~V.~Volotka}
\affiliation{Helmholtz-Institut Jena, Fr\"o{}belstieg 3, D-07743 Jena, Germany}%
\affiliation{Department of Physics, St.~Petersburg State University, 198504 St.~Petersburg, Russia}%

\author{A.~Surzhykov}
\affiliation{Physikalisch-Technische Bundesanstalt, Bundesallee 100, D-38116 Braunschweig, Germany}%
\affiliation{Technische Universit\"at Braunschweig, D-38106 Braunschweig, Germany}%

\author{C.~Z.~Dong}
\affiliation{Key Laboratory of Atomic and Molecular Physics $\&$ Functional Materials of Gansu Province,
             College of Physics and Electronic Engineering, Northwest Normal University, Lanzhou 730070, P.R. China}

\author{S.~Fritzsche}
\affiliation{Helmholtz-Institut Jena, Fr\"o{}belstieg 3, D-07743 Jena, Germany}%
\affiliation{Theoretisch-Physikalisches Institut, Friedrich-Schiller-Universit\"at Jena, Max-Wien-Platz 1, D-07743
Jena, Germany}

\date{\today \\[0.3cm]}

\begin{abstract}

The angular distribution and linear polarization of the fluorescence light following the resonant photoexcitation is
investigated within the framework of the density matrix and second-order perturbation theory. Emphasis has been placed
on ``signatures'' for determining the level sequence and splitting of intermediate (partially) overlapping resonances,
if analyzed as a function of the photon energy of the incident light. Detailed computations within the
multiconfiguration Dirac-Fock method have been performed especially for the $1s^{2}2s^{2}2p^{6}3s\;\, J_{i}=1/2 \,+\,
\gamma_{1} \:\rightarrow\: (1s^{2}2s2p^{6}3s)_{1}3p_{3/2}\;\, J=1/2, \, 3/2
                                                     \:\rightarrow\: 1s^{2}2s^{2}2p^{6}3s\;\, J_{f}=1/2 \,+\, \gamma_{2}$
photoexcitation and subsequent fluorescence emission of atomic sodium. A remarkably strong dependence of the angular
distribution and linear polarization of the $\gamma_{2}$ fluorescence emission is found upon the level sequence and
splitting of the intermediate $(1s^{2}2s2p^{6}3s)_{1}3p_{3/2}\;\, J=1/2, \, 3/2$ overlapping resonances owing to their
finite lifetime (linewidth). We therefore suggest that accurate measurements of the angular distribution and linear
polarization might help identify the sequence and small splittings of closely-spaced energy levels, even if they can
not be spectroscopically resolved.
\end{abstract}

\newpage

 \maketitle

\section{Introduction}
\label{Sec.Introduction}

In atoms and ions with complex shell structures, levels are often closely spaced in energy and, thus, difficult to
resolve spectroscopically. Up to the present, therefore, suitable spectroscopic schemes for resolving the level
structure have played important role in studying the structure of atomic systems
\cite{Martinson/PR:1974,Sonntag/RPP:1992,Vogel/PR:2010}. Experimentally, indeed, great effort has been made to improve
the resolution of photon detectors \cite{Hell/OC:1992,Mueller/NIMPRA:2001,Huotari/JSR:2005} and to obtain ever detailed
spectral information. However, when the level splitting becomes comparable with the (natural) width of the transitions,
it becomes inherently difficult to resolve both, the sequence as well as the splitting of the energy levels owing to
their (partial) overlap, even if high-resolution spectroscopy is applied. In this case, an alternative route to
identify the sequence and splitting of energy levels becomes highly desirable.

In the past decades, much emphasis in atomic spectroscopy has been placed upon the angle-resolved properties of emitted
light, such as the angular distribution and linear polarization \cite{Wu/PRA:2011,Wu/PRA:2012}. When compared to the
total cross sections and decay rates for the photon emission from atoms and ions, angle-resolved measurements were
found more sensitive with regard to the details in the electron-electron and electron-photon interactions. For example,
the angular distribution and linear polarization of fluorescence emission were discussed in studying the Breit
interaction in dielectronic recombination processes
\cite{Fritzsche/PRL:2009,Fritzsche/NJP:2012,Hu/PRL:2012,Joerg/PRA:2015}, the hyperfine interaction
\cite{Surzhykov/PRA-87:2013,Wu/PRA-89:2014} in electron-atom collisions as well as the multipole mixing in the
interaction of ions with the radiation field \cite{Surzhykov/PRL:2002,Weber/PRL:2010}.

However, less attention has been paid to the influence of \textit{overlapping} resonances upon the atomic fluorescence,
in contrast to the autoionization of inner-shell excited atoms \cite{Shimizu/JPB:2000,Ueda/JPB:2001,Kitajima/JPB:2001}.
Only rather recently, we explored the angular and polarization properties of the emitted photons in the two-step
radiative cascade $1s2p^{2}\;\, J_{i}=1/2,\, 3/2 \:\rightarrow\: 1s2s2p\;\, J=1/2,\, 3/2 \,+\, \gamma_1 \:\rightarrow\:
1s^{2}2s\;\, J_{f}=1/2 \,+\, \gamma_1 \,+\, \gamma_2$ of lithium-like tungsten, which proceeds via such overlapping
intermediate resonances \cite{Wu/PRA:2014,Wu/JPCS:2015}. While, for an initially aligned $1s2p^{2}\;\, J_{i}=3/2$
level, a remarkably strong dependence was obtained for the second-step fluorescence photons upon the splitting of the
two (overlapping) $1s2s2p\;\, J=1/2,\, 3/2$ resonances, no effect was found with regard to the sequence of these
resonances due to the mutual cancelation of the sequence-dependent summation terms. In this work, we therefore study
the process of resonant photoexcitation and subsequent fluorescence of atoms to better understand how both, the level
sequence and splitting can be made \textit{visible} for closely-spaced energy levels. To this end, second-order
perturbation theory and the density matrix formalism are employed in order to derive and analyze general expressions
for the angular distribution and linear polarization of the emitted radiation. Though these expressions are applicable
to many-electron atoms (or ions), and are independent of their particular shell structure, we shall consider below the
$2s \:\rightarrow\: 3p$ inner-shell photoexcitation and subsequent fluorescence emissions of sodium atom, e.g.,
$1s^{2}2s^{2}2p^{6}3s\;\, J_{i}=1/2 \,+\, \gamma_{1} \:\rightarrow\: (1s^{2}2s2p^{6}3s)_{1}3p_{3/2}\;\, J=1/2, \, 3/2
\:\rightarrow\: 1s^{2}2s^{2}2p^{6}3s\;\, J_{f}=1/2 \,+\, \gamma_{2}$. For inner-shell excited sodium, the
$(1s^{2}2s2p^{6}3s)_{1}3p_{3/2}\;\, J=1/2, \, 3/2$ resonances are well isolated from other (fine-structure) levels of
the $2s \:\rightarrow\: 3p$ excitation and their level splitting is comparable to the (natural) widths in the
excitation and decay of these resonances.

This paper is structured as follows. In the next section, we present general expressions for the (second-order)
transition amplitude of the photoexcitation and associated radiative decay of the atoms (or ions). This transition
amplitude is then employed to express the angular distribution and linear polarization of the emitted fluorescence
photons. In Sec.~\ref{Sec.ResultsAndDiscussion}, we apply these expressions particularly to the $2s \:\rightarrow\: 3p$
photoexcitation of the $2s^{-1} 3p\;\, ^2P_{1/2,3/2}$ levels in neutral sodium and its subsequent radiative decay back
to the $3s\;\, ^2S_{1/2}$ ground level. Moreover, we later discuss the anisotropy parameter (angular distribution) and
linear polarization of the fluorescence $\gamma_{2}$ photon as functions of incident photon energy, i.e., if the
incident radiation is tuned over the $2s^{-1} 3p\;\, ^2P_{1/2,3/2}$ overlapping resonances. Finally, conclusions and a
brief outlook of the present work are given in Sec.~\ref{Sec.SummaryAndOutlook}.

Atomic units ($m_{e}=1$, $e=1$, $\hbar=1$) are used throughout this paper unless stated otherwise.

\section{Theory and computation}
\label{Sec.TheoryAndComputation}

\begin{figure}[htbp]
\includegraphics[width=8.5cm]{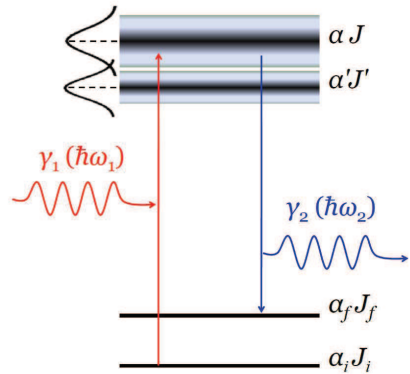}
\caption{\label{FigGenaralTransition} (Color online) Level scheme for the (combined) process of the photoexcitation and
subsequent fluorescence emission via two overlapping resonances. An atom (or ion) in the initial ground level
$\alpha_{i}J_{i}$ absorbs the photon $\gamma_{1}$ and is excited to the overlapping $\alpha J$ and $\alpha^{\prime}
J^{\prime}$ resonances, and subsequently decays to some low-lying levels $\alpha_{f}J_{f}$ via fluorescence emission of
photon $\gamma_{2}$.}
\end{figure}

We here consider the (combined) process
\begin{eqnarray}
\label{GeneralProcess}
   A~(\alpha_{i} J_{i}) \:+\: \gamma_{1}
   & \longrightarrow &
   \left\{ \begin{array}{c}
              A^{*}(\alpha J)   \\[0.1cm]
              A^{*}(\alpha^{\prime} J^{\prime})
           \end{array} \right\}
   \nonumber \\[0.2cm]
   & \longrightarrow &  A(\alpha_{f}J_{f}) \:+\: \gamma_{2} \,
\end{eqnarray}
of the photoexcitation of an atom or ion and its subsequent fluorescence emission, which proceeds via two overlapping
resonances. In contrast to the typical \textit{two-step} model for the excitation and decay, we treat the whole process
(\ref{GeneralProcess}) together in order to allow for a \textit{coherence transfer} during the excitation and decay of
the atoms. In this process, the atom (or ion) is initially assumed to be in its ground level $\alpha_{i} J_{i}$ and is
excited to some overlapping resonances $\alpha J$ and $\alpha^{\prime} J^{\prime}$ by absorbing the photon $\gamma_{1}$
with energy $\omega_{1}$. Owing to the finite lifetime of these resonances, which causes their overlap, they
subsequently decay by photon emission $\gamma_{2}$ (with energy $\omega_{2}$) to some energetically lower-lying levels,
say, $\alpha_{f}J_{f}$; cf. Fig.~\ref{FigGenaralTransition}. While the $J$'$s$ here just denote the total angular
momenta of the levels, the $\alpha$'$s$ refer to all further quantum numbers that are needed for a unique specification
of these levels. In process (\ref{GeneralProcess}), the initial and final levels, $\alpha_{i}J_{i}$ and
$\alpha_{f}J_{f}$, can both be the same, giving rise to the same photon energy $\omega_{1}=\omega_{2}\equiv\omega$ as
we consider in our example below. Let us note here, moreover, that the ``overlap" of the two resonances above is often
caused by fast autoionization channels, and that the (second-step) fluorescence might be suppressed when compared to
the ionization of the system.

\subsection{Evaluation of transition amplitude}
\label{Subsec.TransitionAmplitude}

\begin{figure}[t]
\includegraphics[width=8.5cm]{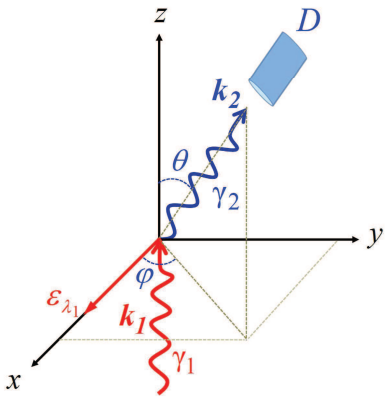}
\caption{\label{FigGeometry} (Color online) Geometry of the photoexcitation and subsequent radiative decay. The
incident light $\gamma_{1}$ propagates along the $z$-axis (chosen as quantization axis) with its polarization vector
$\epsilon_{\lambda_{1}}$ in the $x$-axis, while the fluorescence photon $\gamma_{2}$ is described by the two angles
$(\theta, \varphi)$.}
\end{figure}

The considered (two-step) process (\ref{GeneralProcess}) of the photoexcitation and subsequent fluorescence emission of
an atom or ion is quite analogue to the \textit{resonant} Rayleigh scattering of photons by some atomic target. For the
resonant excitation of atoms ($\omega=E_{\alpha J}-E_{\alpha_{i} J_{i}}$ or $\omega=E_{\alpha' J'}-E_{\alpha_{i}
J_{i}}$), the (Rayleigh) scattering amplitude indeed contains singularities which can be removed by performing an
infinite resummation of the radiative corrections for the resonant levels $\alpha J$ and $\alpha' J'$
\cite{Shabaev/PR:2002,Andreev/PR:2008}. This resummation naturally leads to the occurrence of the linewidths in the
denominators of the second-order transition amplitude. For the case of just two overlapping resonances $\alpha J$ and
$\alpha' J'$, the scattering amplitude in the resonance approximation then takes the form
\begin{widetext}
\begin{eqnarray}
\label{GeneralTransitionAmplitude}
  \mathcal{M}^{\lambda_{1}, \lambda_{2}}_{M_{i}, M_{f}} (\omega)
       & = & \sum_{M'} \frac{ \mem{\alpha_{f} J_{f} M_{f}}
                    {\sum_{m} \bm{\alpha}_{m} \cdot \bm{\epsilon}_{\lambda_{2}}^{\ast} \, e^{-i \bm{k}_{2} \cdot \bm{r}_{m}}}
                    {\alpha' J' M'} \,
                \mem{\alpha' J' M'}
                    {\sum_{m} \bm{\alpha}_{m} \cdot \bm{\epsilon}_{\lambda_{1}} \, e^{i \bm{k}_{1} \cdot \bm{r}_{m}}}
                    {\alpha_{i} J_{i} M_{i}} }
              { E_{\alpha_{i} J_{i}} - E_{\alpha' J'} + \omega + i \, \Gamma_{\alpha' J'} / 2 }
          \nonumber \\[0.2cm]
          &  & + \;\;
         \sum_{M} \frac{ \mem{\alpha_{f} J_{f} M_{f}}
                    {\sum_{m} \bm{\alpha}_{m} \cdot \bm{\epsilon}_{\lambda_{2}}^{\ast} \, e^{-i \bm{k}_{2} \cdot \bm{r}_{m}}}
                    {\alpha J M} \,
                \mem{\alpha J M}
                    {\sum_{m} \bm{\alpha}_{m} \cdot \bm{\epsilon}_{\lambda_{1}} \, e^{i \bm{k}_{1} \cdot \bm{r}_{m}}}
                    {\alpha_{i} J_{i} M_{i}} }
              { E_{\alpha_{i} J_{i}} - E_{\alpha J} + \omega + i \, \Gamma_{\alpha J} / 2 } \, , \hspace{0.08cm}
\end{eqnarray}
where $\bm{k}_{1,2}$ and $\bm{\epsilon}_{\lambda_{1,2}}$ are the wave and polarization vectors of the photons
$\gamma_{1}$ and $\gamma_{2}$, respectively. $\bm{r}_{m}$ and $\bm{\alpha}_{m} \,=\, (\alpha_{x,m}, \alpha_{y,m},
\alpha_{z,m})$ represent the coordinate and the vector of the Dirac matrices for the \emph{m}th electron.
$\ketm{\alpha_{i} J_{i} M_{i}}$ and $\ketm{\alpha_{f} J_{f} M_{f}}$ characterize the initial and final states of the
atom, while the $J$'$s$ and $M$'$s$ refer to the total angular momenta and their projection upon the $z$-axis.
Moreover, $E_{\alpha J} - E_{\alpha_{i} J_{i}}$ and $\Gamma_{\alpha J}$ denote the excitation energy and natural
linewidth of the resonance $\alpha J$, and analogue for the second ``primed" resonance $\alpha' J'$. The operator
$\sum_{m} \bm{\alpha}_{m} \cdot \bm{\epsilon}_{\lambda} \, e^{i \bm{k} \cdot \bm{r}_{m}}$ describes as usual the
interaction of atomic electrons with the radiation field within the velocity gauge in terms of a sum of one-electron
interaction operators. The second-order amplitude (\ref{GeneralTransitionAmplitude}) can be further simplified if the
operator $\bm{\alpha_{m}} \cdot \bm{\epsilon}_{\lambda} \, e^{i \bm{k} \cdot \bm{r_{m}}}$ is decomposed into partial
waves,
\begin{eqnarray}
\label{One-particleEMoperator}
  \bm{\alpha_{m}} \cdot \bm{\epsilon}_{\lambda} \, e^{i \bm{k} \cdot \bm{r_{m}}}
  & = & 4\pi \, \sum_{pLM_{L}} i^{L-p} \, [\bm{\epsilon}_{\lambda} \cdot \bm{Y}^{ (p) \, *}_{LM_{L}} (\hat{\bm{k}})] \;
                                       \bm{\alpha_{m}} \, \bm{a}^{p}_{LM_{L}} (\bm{r_{m}}) \, ,
\end{eqnarray}
where $\bm{Y}^{(p)}_{LM_{L}} (\hat{\bm{k}})$ is a vector spherical harmonics as function of $\hat{\bm{k}} \equiv
\bm{k}/|\bm{k}|$ \cite{Varshalovich:1988} and $\bm{a}^{p}_{LM_{L}} (\bm{r})$ represents the electric ($p=1$) and
magnetic ($p=0$) multipole components of the radiation field. The explicit form of these components has been discussed
at several places elsewhere in the literature \cite{Varshalovich:1988}.

In describing the process (\ref{GeneralProcess}), we here choose the propagation direction $\hat{\bm{k}}_{1}$ of the
incoming photon $\gamma_{1}$ as quantization axis ($z$-axis) and its polarization vector $\bm{\epsilon}_{\lambda_{1}}$
as $x$-axis. Then, the emitted fluorescence photon $\gamma_{2}$ is observed along some direction $\hat{\bm{k}}_{2}$
that is usually characterized by two angles $\hat{\bm{k}}_{2} \equiv (\theta, \varphi)$, the polar angle $\theta$ and
the azimuthal angle $\varphi$ with regard to the $xz$ plane (cf.~Fig.~\ref{FigGeometry}). For this choice of
coordinates, the transition amplitude (\ref{GeneralTransitionAmplitude}) can be written explicitly as
\begin{eqnarray}
\label{SpecificTransitionAmplitude}
  \mathcal{M}^{\lambda_{1}, \lambda_{2}}_{M_{i}, M_{f}} (\omega)
       & = & \sum_{p_{1} L_{1} M_{L_{1}}} \sum_{p_{2} L_{2} M_{L_{2}}} \, i^{L_{1}-L_{2}} \, (i\lambda_{1})^{p_{1}} \,
          (i\lambda_{2})^{p_{2}} \, [L_{1}, L_{2}]^{1/2} \, \delta_{_{\lambda_{1} M_{L_{1}}}} \,
          d^{L_{2}}_{M_{L_{2}}\lambda_{2}} (\theta) \, e^{-i M_{L_{2}} \varphi} \, (-1)^{J_{i}-J_{f}-M_{L_{1}}+1} \,
          \nonumber \\[0.2cm]
       & & \hspace*{-1.3cm} \times \,
          \Bigg\{ \bigg( \sum_{M'} \,
          \sprm{J_{f} M_{f}, L_{2} M_{L_{2}}}{J' M'} \,
          \sprm{J' M', L_{1} -M_{L_{1}}}{J_{i} M_{i}} \bigg) \,
          [J', J_{i}]^{-1/2} \,
          \frac{\mathcal{T}'_{p_{2}L_{2}} \, \mathcal{T}'_{p_{1}L_{1}}}
               { E_{\alpha_{i} J_{i}} - E_{\alpha' J'} + \omega + i \, \Gamma_{\alpha' J'} / 2 } \hspace{0.15cm}
          \nonumber \\[0.2cm]
       & & \hspace*{-1.3cm}  + \,
          \bigg( \sum_{M} \,
          \sprm{J_{f} M_{f}, L_{2} M_{L_{2}}}{J M} \,
          \sprm{J M, L_{1} -M_{L_{1}}}{J_{i} M_{i}} \bigg) \,
          [J, J_{i}]^{-1/2} \,
          \frac{\mathcal{T}_{p_{2}L_{2}} \, \mathcal{T}_{p_{1}L_{1}}}
               { E_{\alpha_{i} J_{i}} - E_{\alpha J} + \omega + i \, \Gamma_{\alpha J} / 2 } \Bigg\} \, .
          \hspace{0.4cm}
\end{eqnarray}
Here, the short-hand notations
$\mathcal{T}_{p_{1}L_{1}}$ $\equiv$ $\redmem{\alpha J} {\sum_{m} \bm{\alpha}_{m} \, \bm{a}^{p_{1}}_{L_{1}}
 (\bm{r}_{m})} {\alpha_{i} J_{i}}$,
$\mathcal{T}'_{p_{1}L_{1}}$ $\equiv$ $\redmem{\alpha' J'} {\sum_{m} \bm{\alpha}_{m} \, \bm{a}^{p_{1}}_{L_{1}}
 (\bm{r}_{m})} {\alpha_{i} J_{i}}$, and
$\mathcal{T}_{p_{2}L_{2}}$ $\equiv$ $\redmem{\alpha_{f} J_{f}} {\sum_{m} \bm{\alpha}_{m} \, \bm{a}^{p_{2}}_{L_{2}}
 (\bm{r}_{m})} {\alpha J}$,
$\mathcal{T}'_{p_{2}L_{2}}$ $\equiv$ $\redmem{\alpha_{f} J_{f}} {\sum_{m} \bm{\alpha}_{m} \, \bm{a}^{p_{2}}_{L_{2}}
 (\bm{r}_{m})} {\alpha' J'}$ are used to denote the reduced transition amplitudes for the absorption of the exciting
photon $\gamma_{1}$ and the emission of the fluorescence photon $\gamma_{2}$, respectively. Moreover, $[a, b] \equiv
(2a+1)(2b+1)$, and the standard notations for the Wigner (small) $d$-function and the Clebsch-Gordan coefficients are
employed.

\subsection{Density matrix of the fluorescence photon}
\label{Subsec.DMofFluorescencePhoton}

Since the transition amplitude (\ref{SpecificTransitionAmplitude}) combines the excitation and the subsequent
fluorescence emission, i.e.\ the photons $\gamma_{1}$ and $\gamma_{2}$, we can quite easily obtain the density matrix
of the fluorescence photon $\gamma_{2}$ from these amplitudes \cite{Balashov:2000,Kabachnik/PR:2007}. For the given
choice of the coordinates, in particular, the density matrix of the photon $\gamma_{2}$ can be expressed in terms of
(the helicity part of) the density matrix of the photon $\gamma_{1}$, sometimes called the helicity density matrix,
\begin{eqnarray}
\label{DMofEmittedPhoton}
  \mem{\hat{\bm{k}}_{2}, \lambda_{2}}{\rho_{\gamma_{2}}}{\hat{\bm{k}}_{2}, \lambda_{2}^{'}}
    & = & \frac{1}{2J_{i}+1} \, \sum_{M_{i}, M_{f}} \,  \sum_{\lambda_{1} \lambda_{1}^{'}} \,
          \mathcal{M}^{\lambda_{1}, \lambda_{2}}_{M_{i}, M_{f}} (\omega) \,
          \mathcal{M}^{\lambda_{1}^{'}, \lambda_{2}^{' \, *}}_{M_{i}, M_{f}} (\omega) \,
          \mem{\hat{\bm{k}}_1, \lambda_{1}}{\rho_{\gamma_{1}}}{\hat{\bm{k}}_1, \lambda_{1}^{'}} \, .
\end{eqnarray}
In deriving this formula, we have assumed that the atom is initially unpolarized and that its final state
$\ketm{\alpha_{f} J_{f} M_{f}}$ remains unobserved in the measurements. In the density matrix theory, moreover, the
helicity density matrix of a photon is a $2\times2$ matrix that just describes polarization of the photon and, is
usually parametrized by means of the three Stokes parameters \cite{Blum:1981,Balashov:2000}
\begin{eqnarray}
\label{Stokes-parameter}
   \mem{\hat{\bm{k}}, \lambda}{\, \rho_{\gamma} \,}{\hat{\bm{k}}, \lambda^{\prime}}
      = \frac{1}{2} \twobytwo{1+P_{3}}{P_{1}-iP_{2}}{P_{1}+iP_{2}}{1-P_{3}} \, .
\end{eqnarray}
Here, $P_{1,2}$ and $P_{3}$ characterize the linear and circular polarization of the photon, respectively. For
unpolarized incident light ($P_1 = P_2 = P_3 = 0$), which is the case that we are just considering in the present work,
in addition, Eq.~(\ref{DMofEmittedPhoton}) can be further simplified to
\begin{eqnarray}
\label{SimplifiedDMofEmittedPhoton}
  \mem{\hat{\bm{k}}_{2}, \lambda_{2}}{\rho_{\gamma_{2}}}{\hat{\bm{k}}_{2}, \lambda_{2}^{'}}
    & = & \frac{1}{2(2J_{i}+1)} \, \sum_{M_{i}, M_{f}} \, \bigg(
    \mathcal{M}^{1,  \lambda_{2}}_{M_{i}, M_{f}} (\omega) \,
    \mathcal{M}^{1,  \lambda_{2}^{' \, *}}_{M_{i}, M_{f}} (\omega) \, + \,
    \mathcal{M}^{-1, \lambda_{2}}_{M_{i}, M_{f}} (\omega) \,
    \mathcal{M}^{-1, \lambda_{2}^{' \, *}}_{M_{i}, M_{f}} (\omega) \, \bigg) \, .
\end{eqnarray}
Since both, the angular distribution and the (linear and circular) polarization of a photon are fully characterized by
its density matrix \cite{Balashov:2000,Kabachnik/PR:2007,Blum:1981}, we are ready now to analyze and discuss these
properties especially for the subsequent fluorescence emission, $\gamma_{2}$.

\subsection{Angular distribution and polarization parameters}
\label{Subsec.ADandLPofFluorescencePhoton}

If, for example, the polarization of the fluorescence photon $\gamma_{2}$ remains unobserved, its angular distribution
simply follows from the trace of the density matrix (\ref{SimplifiedDMofEmittedPhoton}),
\begin{eqnarray}
\label{ADofEmittedPhoton}
  \sigma(\hat{\bm{k}}_{2})
    & = & \mem{\hat{\bm{k}}_{2}, \lambda_{2}=+1
                        }{\rho_{\gamma_{2}}}{\hat{\bm{k}}_{2}, \lambda_{2}^{\prime}=+1} \, + \,
                     \mem{\hat{\bm{k}}_{2}, \lambda_{2}=-1
            }{\rho_{\gamma_{2}}}{\hat{\bm{k}}_{2}, \lambda_{2}^{\prime}=-1} \, .
\end{eqnarray}
For an initially unpolarized atomic target and unpolarized incident photon $\gamma_{1}$, the angular distribution
(\ref{ADofEmittedPhoton}) of the $\gamma_{2}$ fluorescence light is azimuthally symmetric, thus independent of the
angle $\varphi$, and can be characterized by just a single anisotropy parameter $\beta$ if the light is produced by an
electric-dipole (E1) line emission \cite{Balashov:2000},
\begin{eqnarray}
\label{ParametrizedADofEmittedPhoton}
  \rm \sigma(\theta)
    = \frac{\sigma_{0}}{4\pi} \, \Big[ \, 1 + \beta \, P_{2}(\cos\theta) \, \Big] \, .
\end{eqnarray}
It is necessary to mention that this expression is obtained within the E1 approximation. In expression
(\ref{ParametrizedADofEmittedPhoton}), moreover, $\sigma_{0}$ denotes the total scattering cross section and $\rm
P_{2}(\cos\theta)$ is the second-order Legendre polynomial as function of the polar angle $\theta$, taken with regard
to the $z$ axis. Therefore, once we have the anisotropy parameter $\beta$, we also know the angular distribution of the
fluorescence light $\gamma_{2}$.

Apart from the angular distribution, we can use the density matrix (\ref{SimplifiedDMofEmittedPhoton}) to also derive
the linear polarization of the fluorescence radiation. As usual in atomic and optical physics, the linear polarization
is characterized by the two Stokes parameters $P_{1}$ and $P_{2}$ \cite{Blum:1981,Balashov:2000}. For example, the
parameter $P_{1}=(I_{0^{\circ}}-I_{90^{\circ}})/(I_{0^{\circ}}+I_{90^{\circ}})$ is just determined by the intensities
of the fluorescence $\gamma_{2}$ light linearly polarized in parallel ($I_{0^{\circ}}$) or perpendicular
($I_{90^{\circ}}$) with regard to the plane spanned by the propagation direction of the $\gamma_{1}$ and $\gamma_{2}$
photons (cf.~Fig.~\ref{FigGeometry}). Of course, as discussed above, this parameter $P_{1}$ can also be expressed in
terms of the density matrix of the photon by using Eqs.~(\ref{Stokes-parameter})-(\ref{SimplifiedDMofEmittedPhoton}),
\begin{eqnarray}
\label{Pola.ofEmittedPhoton}
  P_{1}(\hat{\bm{k}}_{2})
      = \frac{\mem{\hat{\bm{k}}_{2}, \lambda_{2}=+1}{\,\rho_{\gamma_{2}}}{\hat{\bm{k}}_{2}, \lambda_{2}^{\prime}=-1} \, + \,
              \mem{\hat{\bm{k}}_{2}, \lambda_{2}=-1}{\,\rho_{\gamma_{2}}}{\hat{\bm{k}}_{2}, \lambda_{2}^{\prime}=+1} }
             {\mem{\hat{\bm{k}}_{2}, \lambda_{2}=+1}{\,\rho_{\gamma_{2}}}{\hat{\bm{k}}_{2}, \lambda_{2}^{\prime}=+1} \, + \,
              \mem{\hat{\bm{k}}_{2}, \lambda_{2}=-1}{\,\rho_{\gamma_{2}}}{\hat{\bm{k}}_{2}, \lambda_{2}^{\prime}=-1} } \, .
\end{eqnarray}
With the use of the expressions (\ref{ParametrizedADofEmittedPhoton}) and (\ref{Pola.ofEmittedPhoton}), we can
therefore readily explore the angular distribution and the linear polarization of the characteristic fluorescence light
in the combined excitation and decay process (\ref{GeneralProcess}). Moreover, the $P_{2}$ parameter is always zero for
the case of unpolarized incident light as considered in this work.
\end{widetext}

\subsection{Computation of the reduced matrix elements}
\label{Subsec.Computation}

It follows directly from above that any further analysis of the $\gamma_{2}$ angular distribution and polarization
requires the computation of the second-order transition amplitude (\ref{SpecificTransitionAmplitude}) and, hence, the
(usual) reduced matrix elements $\mathcal{T}_{pL}$ for single-photon bound-bound transitions
\cite{Andrey/PRA:2006,Fritzsche/PRA:2008}. Since these reduced matrix elements occur very frequently in photoexcitation
and radiative transition studies \cite{Grant/AP:1970,Fritzsche/PRA:2005,Inal/PRA:2005}, they are readily available from
different computer codes \cite{Fritzsche/CPC:2012,Gu/CJP:2008}, and not much need to be said about their detailed
computation. We here applied the multiconfiguration Dirac-Fock (MCDF) method \cite{Grant:2007} and especially the
associated \textsc{Grasp92/2K} code \cite{Parpia/CPC:1996,Joensson/CPC:2007} to compute the energy levels and wave
functions of all the relevant atomic states. In the MCDF method, an atomic state function (ASF) with well-defined
parity $P$, total angular momentum $J$ and its component $M$, is approximated by a linear combination of a set of
configuration state functions (CSF) with the same $PJM$,
\begin{eqnarray}
\label{ASF}
   \psi_{\alpha} (PJM) & = & \sum_{r=1}^{n_{c}} \, c_{r}(\alpha)  \ketm{ \phi_{r}(PJM) } \, .
\end{eqnarray}
Here, $n_{c}$ denotes the number of CSF that are used in order to construct the ASF and, $c_{r}(\alpha)$ refers to the
(so-called) configuration mixing coefficients. The CSF are constructed self-consistently on the basis of the
Dirac-Coulomb Hamiltonian, while the relativistic and quantum-electrodynamical effects are incorporated into the
coefficients $c_{r}(\alpha)$ by diagonalizing the matrix of the Dirac-Coulomb-Breit Hamiltonian in first-order
perturbation theory \cite{Parpia/CPC:1996,Joensson/CPC:2007,Grant:2007,Indelicato/PRA:1990,Fritzsche/CPC:2002}. Once
these energy levels and wave functions are obtained, one can easily apply them to calculate all the required reduced
matrix elements by using, for example, the \textsc{Ratip} code \cite{Fritzsche/CPC:2012}.

\begin{figure}[t]
\includegraphics[width=10.0cm]{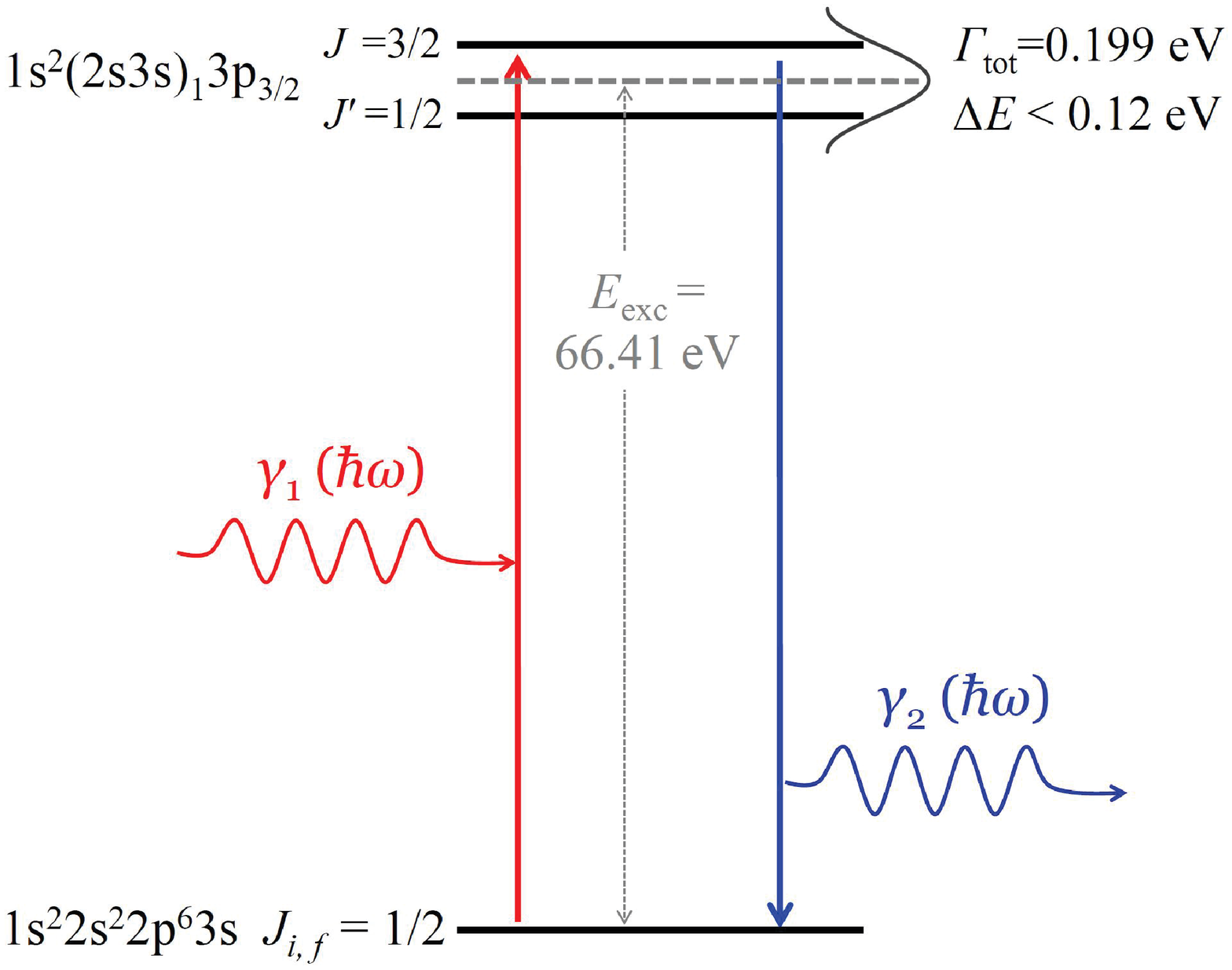}
\caption{\label{FigParticularTransition} (Color online) Level scheme of the $2s \:\rightarrow\: 3p$ inner-shell
photoexcitation (red) and the subsequent radiative decay (blue) of atomic sodium. In the computations, we here use the
experimentally known data for the total linewidth $\Gamma_\mathrm{tot}$ and the central excitation energy $E_{\rm exc}$
\cite{Osawa/JPB:2008}.}
\end{figure}

\section{Results and discussion}
\label{Sec.ResultsAndDiscussion}

\subsection{The $2s \:\rightarrow\: 3p$ photoexcitation and subsequent radiative decay of sodium atom}
\label{Subsec.ParticularCase}

Equations~(\ref{GeneralTransitionAmplitude})-(\ref{SimplifiedDMofEmittedPhoton}) are general and thus applicable to any
atomic (or ionic) system with overlapping (excited) resonances, quite independent of the particular shell structure. As
an example, we shall consider here the $2s \:\rightarrow\: 3p$ photoexcitation of an inner-shell electron and its
subsequent fluorescence emission in atomic sodium,
\begin{eqnarray}
\label{ParticularProcess}
   &  & \hspace*{-2.0cm}
   1s^{2}2s^{2}2p^{6}3s\;\, J_{i}=1/2 \:+\: \gamma_{1}
   \nonumber \\[0.1cm]
   & \longrightarrow & \left\{ \begin{array}{l}
                                  (1s^{2}2s2p^{6}3s)_{1}3p_{3/2}\;\, J =1/2   \\[0.1cm]
                                  (1s^{2}2s2p^{6}3s)_{1}3p_{3/2}\;\, J'=3/2
                               \end{array}
                       \right\}
   \nonumber \\[0.1cm]
   & \longrightarrow &  1s^{2}2s^{2}2p^{6}3s\;\, J_{f}=1/2 \:+\: \gamma_{2} \, .
\end{eqnarray}
Note, that all the transitions are E1-allowed in this excitation and decay scheme (\ref{ParticularProcess}),
cf.~Fig.~\ref{FigParticularTransition}. In the following, we can therefore restrict ourselves to the E1 approximation,
i.e.~to $p_{1}=p_{2}=1$, and $L_{1}=L_{2}=1$ in Eq.~(\ref{SpecificTransitionAmplitude}). In this approximation, just
two reduced E1 matrix elements $\redmem{J_{f}=1/2}{\sum_{m} \bm{\alpha}_{m} \bm{a}^{1}_{1} (\bm{r}_{m})} {J=1/2, 3/2}$
need to be calculated in order to obtain the transition amplitude (\ref{SpecificTransitionAmplitude}), where we have
omitted the electron configurations for the sake of brevity. The two \ket{J=1/2, 3/2} resonances \textit{overlap} each
other and cannot be resolved spectroscopically \cite{LaVilla/JPB:1981,Journel/JP:1993,Juranic/PRA:2006,Osawa/JPB:2008}.
In the computations, we use the experimentally known data 0.199~eV and 66.41~eV for the total linewidth
$\Gamma_\mathrm{tot}$ and the central excitation energy $E_{\rm exc}$ \cite{Osawa/JPB:2008}. Moreover, since the
$(1s^{2}2s2p^{6}3s)_{1}3p_{3/2}\;\, J =1/2, 3/2$ overlapping resonances are well isolated from other excited levels of
neutral sodium \cite{NIST-Database}, we just `tune' the resonances with the incident $\gamma_{1}$ light and omit all
other excitations. In this particular example of sodium, the two resonances remain unresolved spectroscopically for the
level splittings $|\Delta E| \equiv |E_{3/2} - E_{1/2}| \lesssim$ 0.12~eV due to a resolution criteria for two
overlapping resonances with (approximately) the same individual linewidths and the total linewidth $\Gamma_{\rm tot}
\simeq 0.2$~eV. In our analysis below, therefore, we consider also level splittings $|\Delta E|$ which are smaller than
the resolution criteria of 0.12~eV. For other splittings (much) larger than 0.12~eV, they are not physically
significant since the yield of the subsequent fluorescence $\gamma_{2}$ photons is almost null when tuning photon
energy of the incident $\gamma_{1}$ light between the two resonances. While the total linewidth of the overlapping
resonances is experimentally known for neutral sodium, their individual linewidths are estimated to be the same and are
expressed approximately as $\Gamma_{\alpha J} = \Gamma_{\alpha' J'} \simeq \Gamma_{\rm tot} (1 - \Delta
E^{2}/2\Gamma_{\rm tot}^{2})$ in terms of the total linewidth $\Gamma_{\rm tot}$ and the (assumed) level splitting
$\Delta E$.

Below, we shall apply Eqs.~(\ref{GeneralTransitionAmplitude})--(\ref{Pola.ofEmittedPhoton}) in order to analyze the
angular distribution and linear polarization of the $\gamma_{2}$ fluorescence emission following the $2s
\:\rightarrow\: 3p$ photoexcitation of sodium via the $(1s^{2}2s2p^{6}3s)_{1}3p_{3/2}\;\, J =1/2, 3/2$ overlapping
resonances. In particular, we are interested how this fluorescence depends on both, the \textit{level sequence} and the
\textit{splitting} of the $J =1/2, 3/2$ resonances, if analyzed as a function of the photon energy of the incident
light $\gamma_{1}$. In addition, we shall propose two independent scenarios for determining experimentally the sequence
and splitting by measuring the angular distribution and linear polarization of the fluorescence light.

\subsection{Angular distribution of the fluorescence photons}
\label{Subsec.ADofFluorescencePhoton}

\begin{figure}[b]
\includegraphics[width=9.5cm]{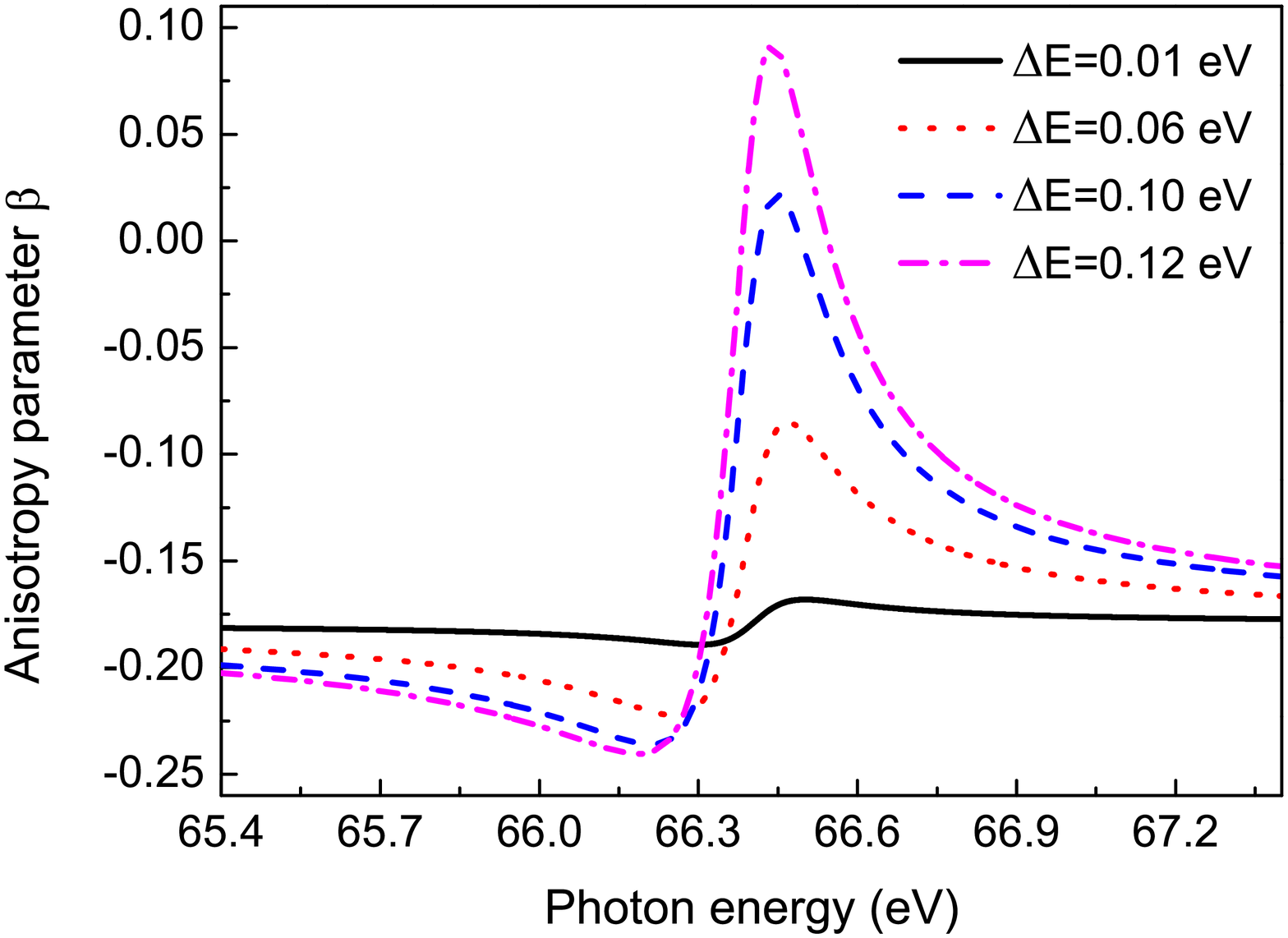}
\caption{\label{FigAnisoPara-DifferentE} (Color online) Anisotropy parameter $\beta$ for the angular distribution of
the $(1s^{2}2s2p^{6}3s)_{1}3p_{3/2}\;\, J=1/2, 3/2 \:\rightarrow\: 1s^{2}2s^{2}2p^{6}3s\;\, J_{f}=1/2$ fluorescence
emission of sodium as functions of the photon energy $\omega$ of the incident $\gamma_{1}$ light. Results are shown for
several assumed level splittings of the two $J=1/2, 3/2$ overlapping resonances: $\triangle E=$ 0.01~eV (black solid
line), 0.06~eV (red dotted line), 0.10~eV (blue dashed line), and 0.12~eV (magenta dash-dotted line).}
\end{figure}

Let us start with the angular distribution of the $\gamma_{2}$ fluorescence emission from the two
$(1s^{2}2s2p^{6}3s)_{1}3p_{3/2}\;\, J=1/2, 3/2$ overlapping resonances after the photoexcitation. For different
$\gamma_{1}$ photon energies of the incident light, the population of these levels is expected to differ relative to
each other and so also the angular distribution of the fluorescence emission. Moreover, the (coherent) excitation of
the two resonances also depends on the level splitting \cite{Wu/PRA:2014,Wu/JPCS:2015} and, this should thus become
visible in the angular distribution as well.

Figure~\ref{FigAnisoPara-DifferentE} displays the anisotropy parameter $\beta$ for the angular distribution of the
$(1s^{2}2s2p^{6}3s)_{1}3p_{3/2}\;\, J=1/2, 3/2 \:\rightarrow\: 1s^{2}2s^{2}2p^{6}3s\;\, J_{f}=1/2$ fluorescence
emission of sodium as functions of the photon energy $\omega$ of the incident light. Results are shown for different
level splittings $|\Delta E| \equiv |E_{3/2} - E_{1/2}| = 0.01$, 0.06, 0.10, and 0.12~eV of the two overlapping
resonances of sodium, which we assumed to be \textit{variable}. As seen from this figure, the anisotropy parameter
appears to be very sensitive for (almost) all level splittings with regard to the photon energy of the incident light.
Moreover, this $\beta$ parameter also strongly depends upon the level splitting itself near the resonances. This latter
dependence arises from the Lorentzian shape of the excitation distribution of the resonances due to their finite
natural width.

\begin{figure}[t]
\includegraphics[width=9.5cm]{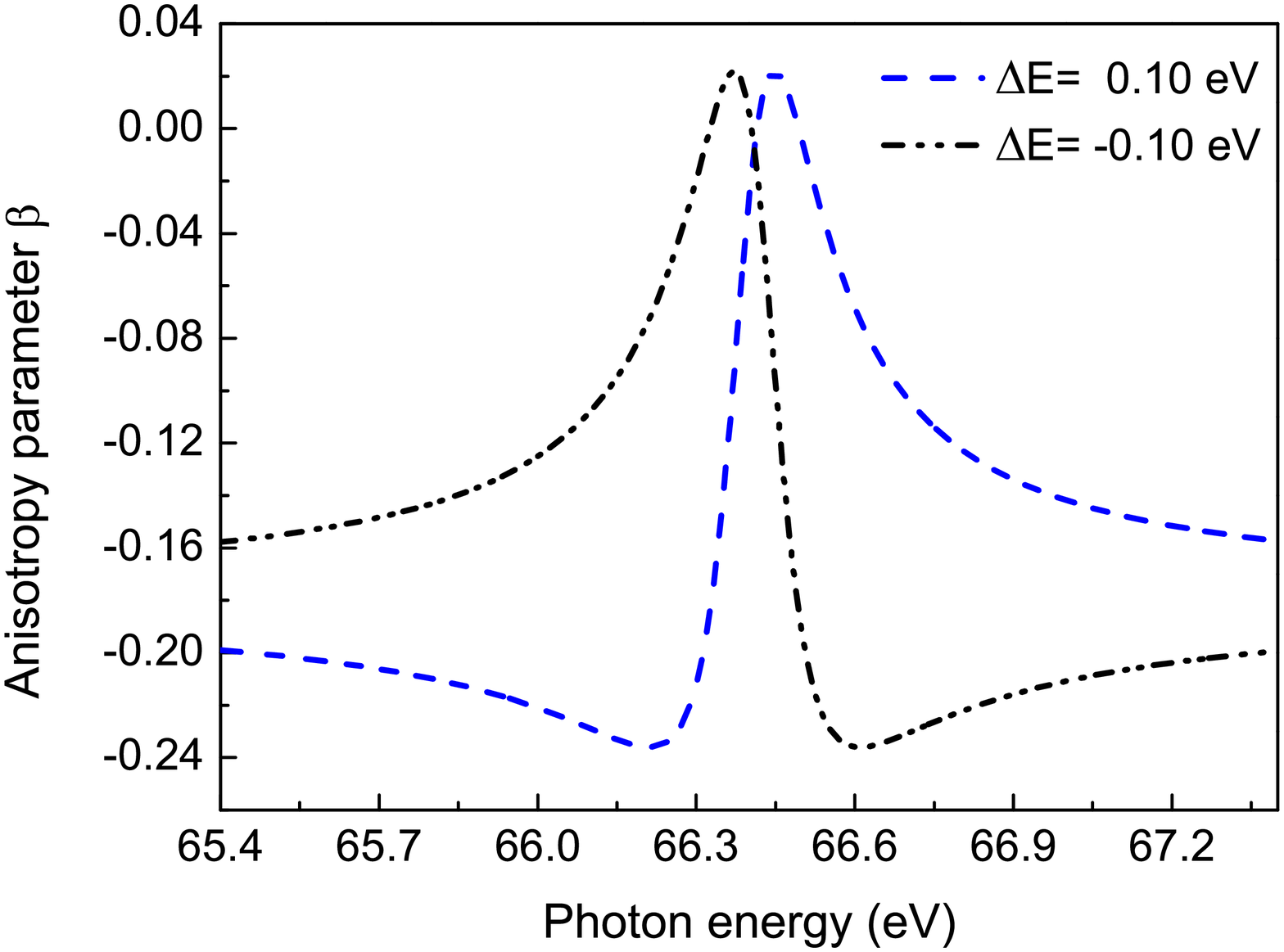}
\caption{\label{FigAnisoPara-OppositeE} (Color online) Anisotropy parameter $\beta$ for the angular distribution of the
$(1s^{2}2s2p^{6}3s)_{1}3p_{3/2}\;\, J=1/2, 3/2 \:\rightarrow\: 1s^{2}2s^{2}2p^{6}3s\;\, J_{f}=1/2$ fluorescence
emission of sodium as functions of the photon energy $\omega$ of the incident $\gamma_{1}$ light. Results are shown for
two assumed level splittings $\triangle E=$ 0.10~eV (blue dashed line) and -0.10~eV (black dash-dot-dotted line), which
indicate that the (absolute) splitting of the two overlapping resonances remains the same but the sequence becomes
opposite in both cases.}
\end{figure}

Apart from the level splitting, the anisotropy parameter $\beta$ of the fluorescence light also depends on the level
\textit{sequence} if the photon energy of the incident light is tuned over the resonances. This can be seen from
Fig.~\ref{FigAnisoPara-OppositeE}, where we plot the $\beta$ parameter as a function of the photon energy but for two
opposite level splittings $\triangle E=$ 0.10~eV and -0.10~eV, respectively. The opposite sign here indicates that the
(absolute) splitting remains the same but that the level sequence has been reversed for the two overlapping resonances.
For instance, the negative level splitting $\triangle E=$ -0.10~eV means that the $J=3/2$ resonance lies lower than the
$J=1/2$ one energetically by 0.10~eV. In particular, the shape of the two $\beta$ distributions occurs to be symmetric
with regard to the central excitation energy of the overlapping resonances. This can be readily understood from
formula~(\ref{GeneralTransitionAmplitude}) since the reversal of the level sequence is equivalent to the interchange in
the $\gamma_{1}$ photon energy with regard to the central energy. This differs from the predicted angular distribution
of the emitted fluorescence light in the two-step radiative cascade of W$^{71+}$ ions, which was found insensitive with
regard to the level sequence of the overlapping resonances due to the mutual cancelation of the sequence-dependent
summation terms \cite{Wu/PRA:2014}. Therefore, accurate angular measurements of the fluorescence emission following the
photoexcitation with sufficiently `thin-banded' incident light might help identify both, the level sequence and
splitting of closely-spaced energy levels in excited atoms or ions.

\subsection{Linear polarization of the fluorescence photons}
\label{Subsec.LPofFluorescencePhoton}

\begin{figure}[b]
\includegraphics[width=9.5cm]{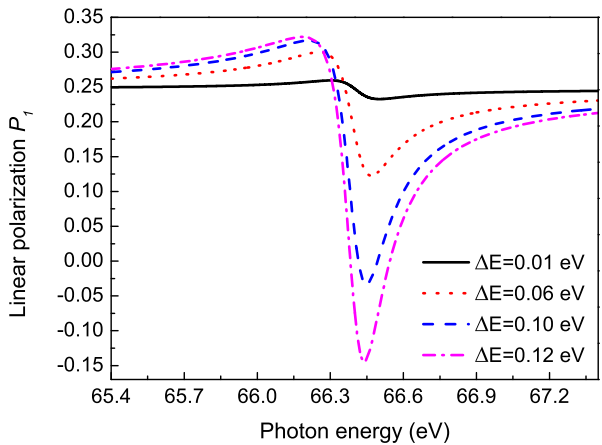}
\caption{\label{FigPolari90P1-DifferentE} (Color online) The same as Fig.~\ref{FigAnisoPara-DifferentE} but for the
linear polarization of the fluorescence $\gamma_{2}$ light emitted perpendicular ($\theta=90^{\circ}$) to the
propagation direction of the incident $\gamma_{1}$ light.}
\end{figure}

Until now, we just have discussed the angular distribution of the fluorescence emission from $2s \rightarrow\, 3p$
photo-excited sodium, if the photon energy of the incident light is tuned over the two
$(1s^{2}2s2p^{6}3s)_{1}\,3p_{3/2}\;\, J=1/2, 3/2$ overlapping resonances at about 66.4~eV. Alternatively, we may
consider and analyze also the linear polarization of this fluorescence light, which can be measured nowadays with quite
high accuracy either by means of solid-state Compton polarimeters \cite{Tashenov/PRL:2006,Tashenov/PRL:2011} or, even
more precisely, with Bragg crystal polarimeters \cite{Beiersdorfer/PRA:1996,Nakamura/PRA:2001,Kaempfer/PRA:2016}.

\begin{figure}[t]
\includegraphics[width=9.5cm]{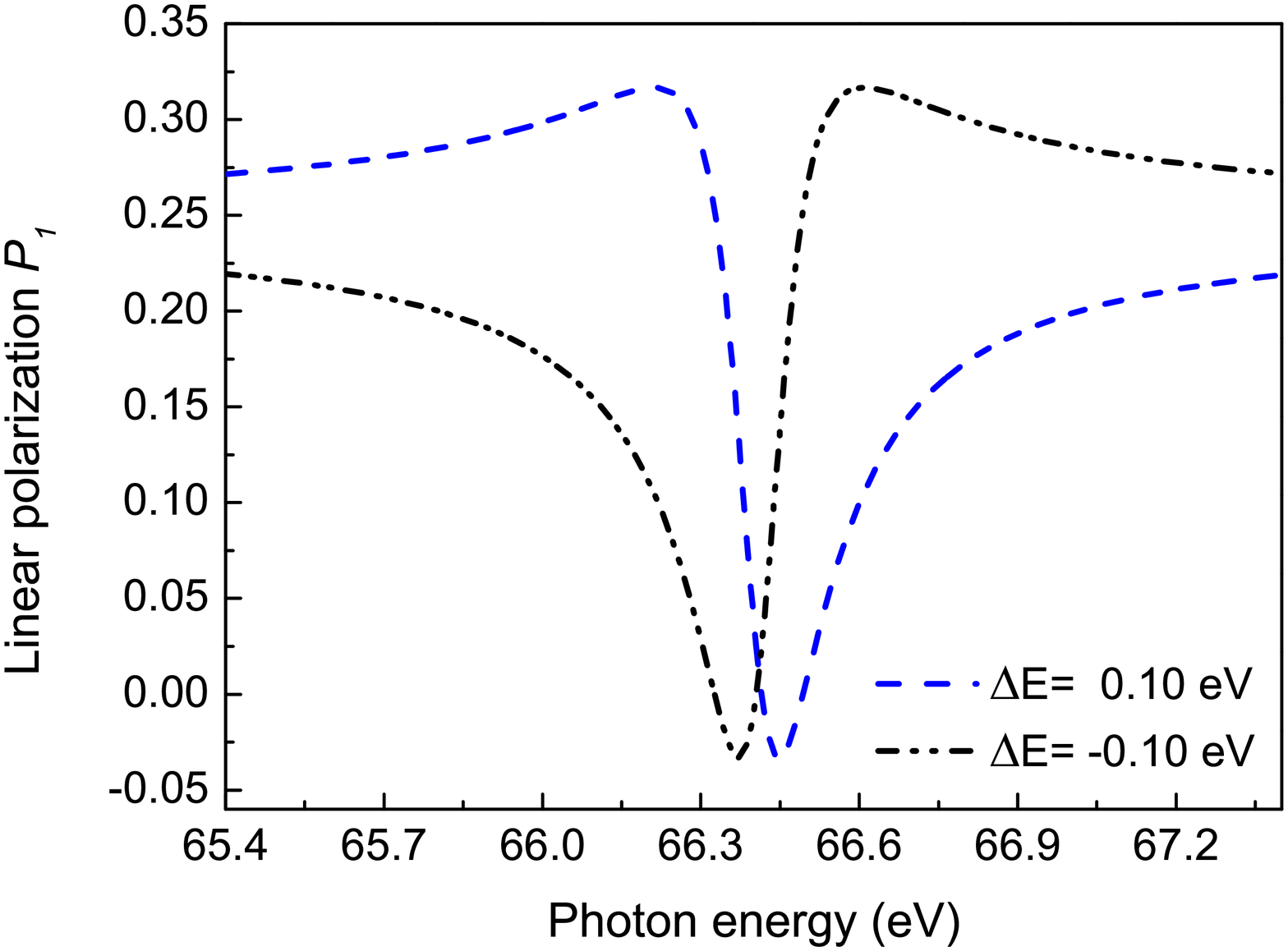}
\caption{\label{FigPolari90P1-OppositeE} (Color online) The same as Fig.~\ref{FigAnisoPara-OppositeE} but for the
linear polarization of the fluorescence $\gamma_{2}$ light emitted perpendicular ($\theta=90^{\circ}$) to the
propagation direction of the incident $\gamma_{1}$ light.}
\end{figure}

In Fig.~\ref{FigPolari90P1-DifferentE}, we therefore display the linear polarization of the fluorescence light
$\gamma_{2}$ that is emitted perpendicular ($\theta=90^{\circ}$) to the propagation direction of the incident light
$\gamma_{1}$. Again, the degree of linear polarization is shown as functions of the frequency of the incident light and
for the same (assumed) level splittings of the two $(1s^{2}2s2p^{6}3s)_{1}\,3p_{3/2}\;\, J=1/2, 3/2$ overlapping
resonances. Similar as for the angular distribution, the degree of linear polarization depends on the level sequence
and splitting, and here even at a larger absolute scale and with a change of its sign near to the central transition
frequency. In addition, Fig.~\ref{FigPolari90P1-OppositeE} displays the linear polarization for a level splitting of
0.1~eV and its symmetry with regard to the central frequency if the level sequence is interchanged. As seen from these
figures, accurate polarization measurements could also serve as an alternative and independent route to identify the
sequence and splitting of overlapping resonances. Finally, we have to mention here that the incident $\gamma_{1}$ light
is assumed to be monochromatic in the angular and polarization analysis of the fluorescence photon above. The use of a
non-monochromatic incident light could weaken (more or less) the obtained angular and polarization dependence upon the
(central) energy of the light, depending on the linewidth of the laser used. Nevertheless, this dependence still
remains strong enough to be observable by using the present-day photon detectors.

\section{Summary and outlook}
\label{Sec.SummaryAndOutlook}

In summary, the photoexcitation and subsequent fluorescence emission of atoms have been studied within the framework of
the density matrix and second-order perturbation theory. Attention has been paid especially to the angular distribution
and the linear polarization of the fluorescence as observed from (partially) overlapping resonances and how such
measurements would be affected by the level sequence and splitting of the resonances involved. Detailed MCDF
calculations were performed for the $1s^{2}2s^{2}2p^{6}3s\;\, J_{i}=1/2 \,+\, \gamma_{1} \:\rightarrow\:
(1s^{2}2s2p^{6}3s)_{1}\,3p_{3/2}\;\, J=1/2, \, 3/2 \:\rightarrow\: 1s^{2}2s^{2}2p^{6}3s\;\, J_{f}=1/2 \,+\, \gamma_{2}$
photoexcitation and subsequent photon emission of sodium atoms. It is predicted that the angular distribution and
linear polarization of the $\gamma_{2}$ fluorescence photons strongly depend upon the level sequence and splitting of
the $(1s^{2}2s2p^{6}3s)_{1}\, 3p_{3/2}\;\, J=1/2, \, 3/2$ resonances, if analyzed as functions of the frequency of the
incident light. This dependence is caused by the non-negligible linewidth of the (overlapping) resonances which lead to
a coherence transfer in the population of the resonances. This coherence transfer also affects the angular and
polarization properties of the emitted fluorescence light. We therefore suggest that accurate measurements of the
angular distribution and linear polarization of fluorescence light can be utilized to help identify the sequence and
splitting of closely-spaced atomic (or ionic) energy levels, even if these levels cannot be resolved spectroscopically.

With the recent progress of light sources and photon detection techniques, the proposed measurements are feasible
today. For example, laser-induced fluorescence spectroscopy or synchrotron radiation can be utilized for such
energy-selective measurements of the subsequent fluorescence emission. In addition, the change of the obtained
anisotropy parameter and linear polarization is large enough as functions of the incident photon energy to be measured
by using present-day photon detector and polarimeter.

\begin{acknowledgments}
Z.W.W. acknowledges the support from the Helmholtz Institute Jena and the Research School of Advanced Photon Science of
Germany. This work has been supported by the BMBF under Contract No. 05K13VHA and by the National Natural Science
Foundation of China under Grant Nos. 11464042, 11274254, and U1332206.
\end{acknowledgments}

\end{document}